\documentclass[journal, onecolumn]{IEEEtran}
\usepackage{amsmath}
\usepackage{graphicx}
\usepackage{adjustbox}  
\usepackage{cuted}
\usepackage{enumitem}

\ifCLASSINFOpdf
\else
\fi
\begin{document}
%
\title{High-Q Interstitial Square Coupled Microring Resonators Arrays}
%
%
%

\author{Shaolin Liao$^{1*}$,~\IEEEmembership{ Senior Member,~IEEE,}
        Lu~Ou$^2$,~\IEEEmembership{Member,~IEEE}
\thanks{$^1$S. Liao is with Department of Electrical and Computer Engineering, Illinois Institute of Technology, Chicago, IL 60616 USA.}
\thanks{$^*$Corresponding author (sliao5@iit.edu).}
\thanks{$^2$L. Ou is with College of Computer Science and Electronic Engineering, Hunan University, Changsha, Hunan, China  410082.}
\thanks{Manuscript received 2020.}}

%
%

\markboth{July, 2020}%
{Shell \MakeLowercase{\textit{et al.}}: Bare Demo of IEEEtran.cls for IEEE Journals}
%



\maketitle

\begin{abstract}
The properties of the square array of coupled Microring Resonators (MRRs) with interstitial rings are studied. Dispersion behavior of the interstitial  square coupled MRRs is obtained through the transfer matrix method with the Floquet-Bloch periodic condition. Analytical formulas of the eigen wave vectors, band gaps and eigen mode vectors are derived for the special cases of the interstitial square coupled MRRs array with identical couplers and the regular square coupled MRRs array without the interstitial rings. Then, the eigen modes' field distribution are calculated for each of the four eigen wave vectors for a given frequency through the secular equation. Finally, numerical simulation is performed for an interstitial square coupled MRRs array with identical couplers and a regular square coupled MRRs array. The simulation result verifies the analytical analysis. Finally, the loaded quality factors of  the interstitial 5-ring configuration, the regular 4-ring configuration and the 1-ring configuration are obtained. It is found that the loaded quality factor of the interstitial 5-ring configuration is up to 20 times and 8 times as high as those of the 1-ring configuration and the regular 4-ring configuration respectively, mainly due to the degenerated eigen modes at the resonant frequency. Thus, the interstitial square coupled MRRs array has the great potential to form high-quality integrated photonics components, including  filters and resonance based sensing devices like the parity-time symmetric sensors.
\end{abstract}

\begin{IEEEkeywords}
Microring resonators (MRRs), periodic array, dispersion, quality factor Q.
\end{IEEEkeywords}

%
\IEEEpeerreviewmaketitle

\section{Introduction}
%
%
%
%
\IEEEPARstart{C}{ompared} to low-frequency electromagnetic waves \cite{Liao_Ping_Pong_APMC_2020}-\cite{Liao_overmode_JEMWA_2008},  optics wave has the advantage of broad bandwidth \cite{Liao_ZIM_arXiv_2020}-\cite{Liao_X_Ray_arXiv_2020}. In particular, photonics integrated circuitry has been critical for optical network and sensing platform due to its compactness, fast speed, broad bandwidth and low-power consumption. On one hand, the next-generation communication network is pushing the optical infrastructure to the last-mile end user for higher data rate due to its broader bandwidth; On the other hand, the demand on higher sensitivity of the sensing platform requires better photonics integrated circuits.  To achieve multi-function photonics integrated circuits, more optical components and devices are required on an optical chip of small footprint.

In the past decades, Microring Resonators (MRRs) have been considered as one of the strong candidates for building blocks of the optical components and devices, due to its microns in size and simple planar structure suitable for fabrication. MRRs are basically close-loop light wave guiding structures that circulate the light wave around it for many times. When the round-trip length is an integer number of wavelengths, resonance forms within the MRRs. The transmission phases of the MRRs at resonance change abruptly and the group velocities reduce dramatically, providing an efficient way to slow the light wave. Such slow-light phenomenon has been explored to realize many critical optical components/devices, including the high-performance filter such as add-drop filters \cite{robinson_photonic_2013} and wavelength division multiplexers \cite{yu_8y8_2013}, \cite{dunmeekaew_new_2010}, optical delay lines \cite{katti_photonic_2018},  Parity-Time (PT) symmetric devices \cite{miri_exceptional_2019}, nonlinear light-wave and materials interaction such as four wave mixing \cite{zeng_four-wave_2015}, optical parametric generation \cite{luo_optical_2019}, \cite{borghi_four_2019}, single-photon/photon-pair source \cite{guo_nonclassical_2019}, \cite{heuck_unidirectional_2019}, \cite{alsing_photon_pair_2017}, frequency comb generation \cite{zhang_broadband_2019}, optical quantum computing \cite{scott_scalable_2019}, as well as high-quality sensors \cite{amiri_high_2018}, \cite{wu_label_free_2019}, \cite{steglich_optical_2019}, \cite{barrios_integrated_2012}.  Depending on its geometry, two typical MRRs are actively studied, {\it i.e.}, the circular MRRs and the racetrack MRRs. The performance of optical components/devices made of MRRs depends on their quality factor $Q$, which is a function of the ratio between the stored energy within the MRRs and the energy loss \cite{malthesh_improvement_2019}, \cite{zhang_design_2018}, \cite{nada_theory_2017}. To achieve optical components/devices of high $Q$, better MRRs structures are needed for efficient energy confinement and storage and loss has to be reduced \cite{malthesh_improvement_2019}.  On one hand, to design better MRRs structures, multi-ring geometries and ring-waveguide coupling schemes have to be designed. The Coupled Resonator Optical Waveguide (CROW \cite{nada_theory_2017}, \cite{poon_designing_2004}, \cite{poon_matrix_2004}, \cite{poon_polymer_2006}) and the Side-Coupled Integrated Space Sequenced Optical Resonator (SCISSOR \cite{e_heebner_slow_2002}, \cite{mancinelli_optical_2011}) are great examples to achieve high-performance filters; On the other hand, to reduce the loss, better materials and fabrication process are needed.

Although microring resonators can be made on platform of various materials \cite{han_simulation_2019}, \cite{ramelow_strong_2019}, \cite{zhang_monolithic_2017}, \cite{zheng_high_quality_2019}, \cite{Krasnokutska_2019}, it is well known that silicon-on-insulator process has been very advanced, thanks to the rapid progress of the silicon electronic applications. So mass production of high-quality scalable MRRs on the silicon platform is made possible \cite{bogaerts_silicon_2012}, \cite{tan_silicon_2018}. Another advantage of silicon platform is its high refractive index contrast: the silicon core has a refractive index of $\sim 3.47$ and the silicon oxide bottom substrate cladding has  a refractive index of $\sim 1.44$. Such high refractive index contrast allows very compact light wave guiding silicon waveguide/wire with width and height down to $\sim 400-500 nm$ and $\sim 200-250 nm$ respectively \cite{bogaerts_silicon_2012}. It has been shown that MRRs with radii down to $1 \mu m$ can be realized \cite{bogaerts_silicon_2012}, \cite{tan_silicon_2018}. 

In the literature, many works have been devoted to the study of single/double/multiple MRRs and 1D arrays, including single waveguide bus coupled to MRRs, double waveguide buses coupled to MRRs, multiple-ring CROW \cite{nada_theory_2017}, \cite{radjenovic_2017},  and SCISSOR light guiding structures \cite{e_heebner_slow_2002}, \cite{mancinelli_optical_2011}. However, 2D array of MRRs is less explored. Due to its importance in realizing high-quality multiple-ring resonant structures \cite{peng_zim_2019}, in this paper, the 2D interstitial squared coupled MRRs array is studied. Specifically, dispersion curves between the light wave propagation vector and its corresponding frequency is obtained through the Transfer Matrix Method (TMM) for the periodic Floquet-Bloch condition \cite{liao_optimal_2019}; Analytical formulas of eigen wave vectors, band gaps and eigen modes are derived for the special cases of identical couplers and the regular square coupled MRRs array without the interstitial ring coupling. Then, numerical simulation is performed to verify the analytical analysis.

\begin{figure}[ht]
 \centering
\includegraphics[scale=0.45]{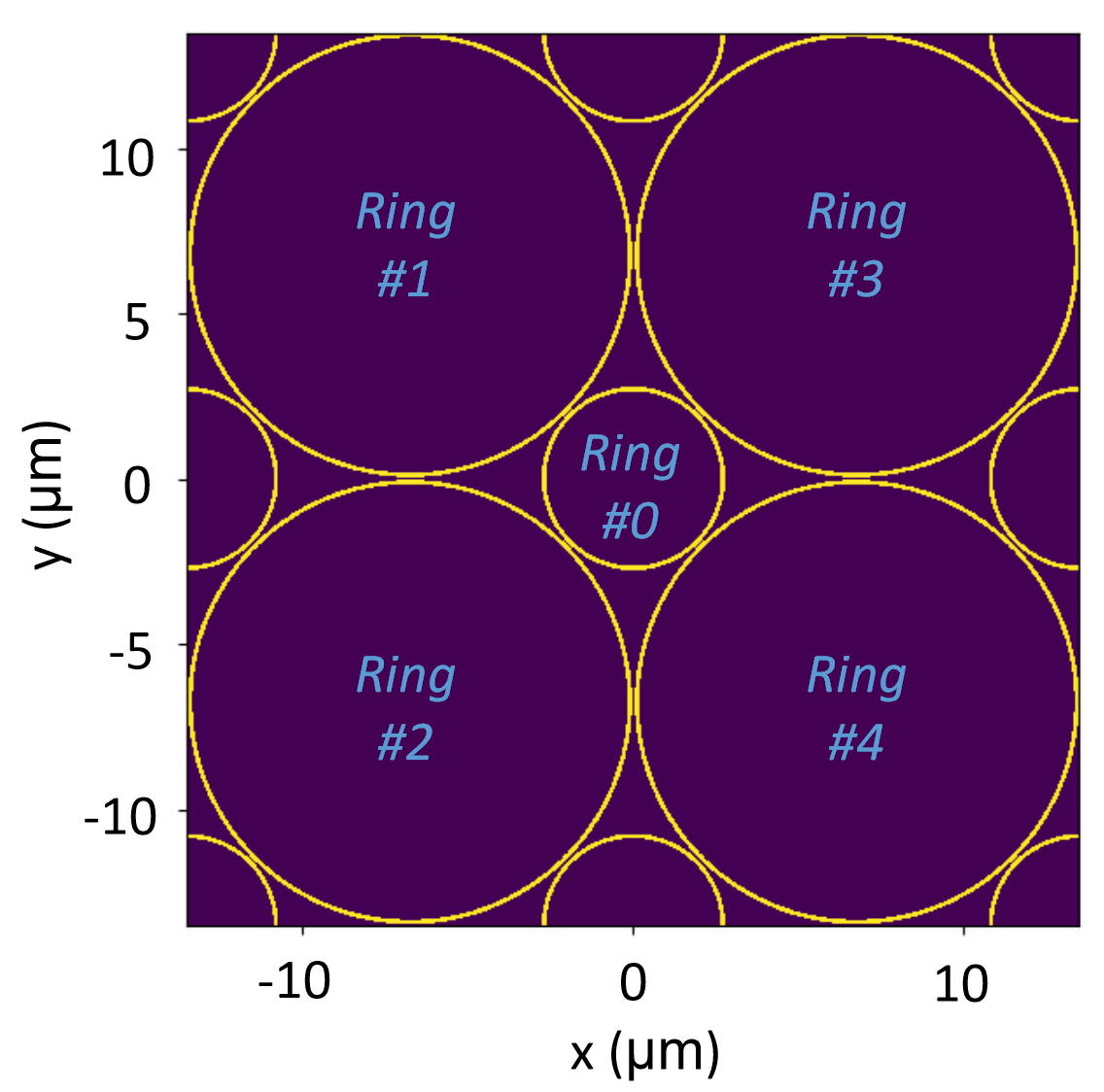}
\caption{\label{fig:4unit_cell} The geometry of  the 2D interstitial square coupled MRRs array showing 4 unit cells.}
\end{figure}

\begin{figure}[ht]
 \centering
\includegraphics[scale=0.45]{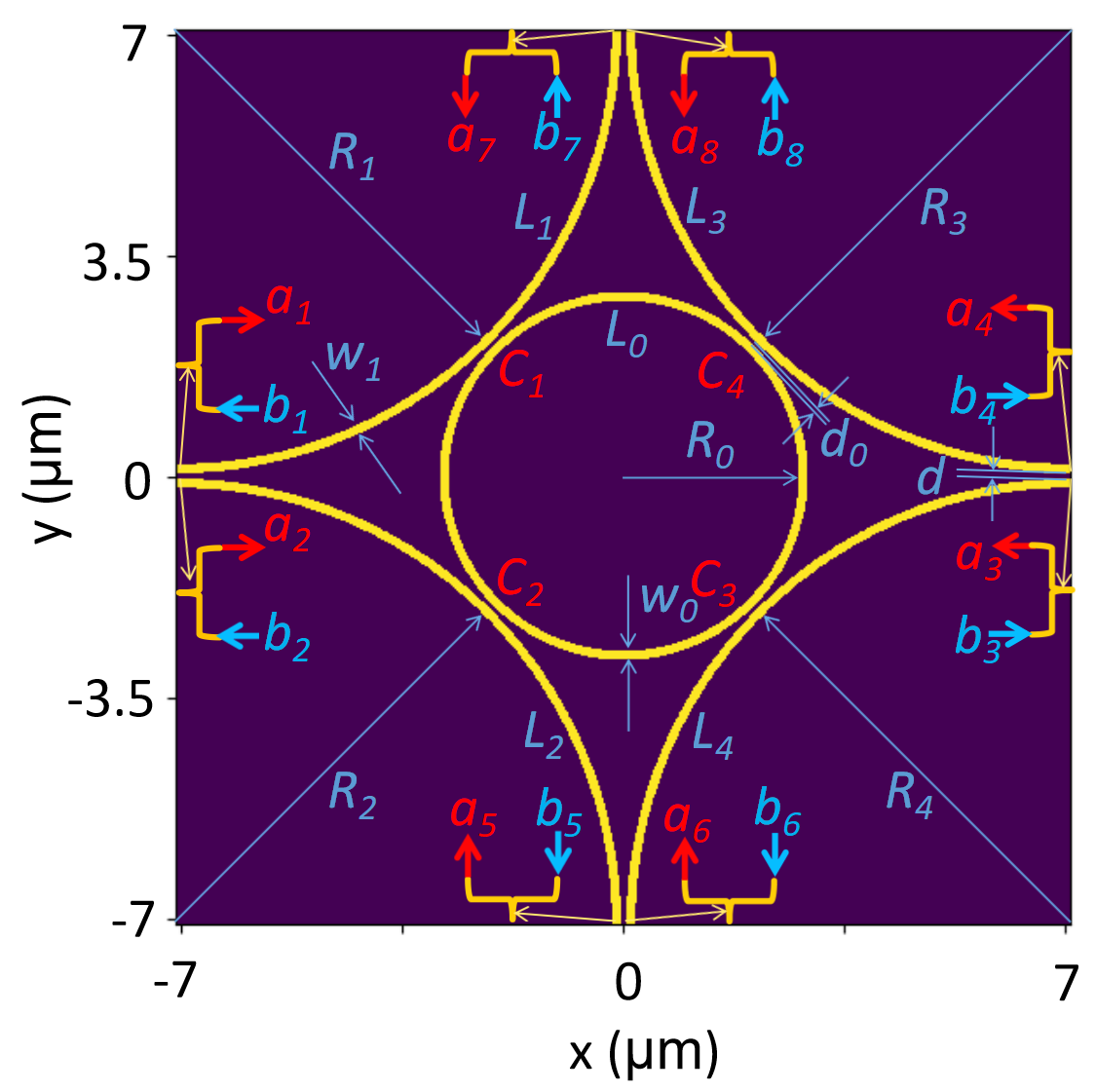}
\caption{\label{fig:unit_cell} One unit cell showing the dimensions.}
\end{figure}

The rest of the paper is organized as follows. First, in Section \ref{sec:array}, the geometry of the 2D interstitial square coupled MRRs array is shown; followed by the TMM formulation of the problem in Section \ref{sec:TMM}, as well as the derivation of transfer matrix of the 2D interstitial coupled MRRs array and the secular equation in Section \ref{sec:TMM_2D}; then the analytical analysis of the dispersion relation, the band gaps and the eigen modes are presented in Section \ref{sec:disp}, Section \ref{sec:band_gap} and Section \ref{sec:eigen_modes} respectively; finally, the numerical simulation and results are shown in Section \ref{sec:results}; followed by the application discussion in Section \ref{sec:application} and the conclusion of the paper in Section \ref{sec:con}.

\section{2D Interstitial Square Coupled MRRs Array} \label{sec:array}

The geometry structure of the 2D interstitial square microring resonators array is shown in Fig. \ref{fig:4unit_cell} where 4 unit cells are shown. Rings $(\#1, \#2, \#3, \#4)$ are the larger site-rings on the 2D square lattice and ring $(\#0)$ is the smaller interstitial ring.  The 4 larger site-rings are identical in size and coupled to the middle smaller interstitial-ring. The unit cell of the 2D interstitial square coupled MRRs array is shown in Fig. \ref{fig:unit_cell}. Also shown are dimensions for the structure: $R_i/R_0, =1, 2, 3, 4$ and $w/w_0$ are the radii and widths  of the larger site-ring and smaller interstitial-ring respectively; similarly, $d/d_0$ are the gap distance between the larger site-rings and the gap distance between the larger site-ring and the smaller interstitial-ring respectively; finally, $L_i, i=1, 2, 3, 4$ are the $45^\circ$-arc lengths of the larger site-rings and $L_0$ is the $45^\circ$-arc length of the smaller  interstitial-ring. Couplers between the site-rings (at the left/bottom/right/top locations) are called the site-site couplers and couplers between the site-ring and the interstitial-ring (at the $C_1/C_2/C_3/C_4$ locations) are called the site-interstitial couplers. The coupling coefficients of the site-site coupler and the site-interstitial coupler are defined as $\kappa_i, i=1, 2, 3, 4$ and $\kappa_0$ respectively \cite{kaplan_modeling_2006}. In particular, the interstitial square coupled MRRs array has the $C_{4v}$ symmetry \cite{sakoda_dirac_2012} when $R_i = R; L_i=L, i=1, 2, 3, 4$ and $\kappa_i = \kappa_0$. 

\section{TMM Formulation}\label{sec:TMM}
For periodic structure, Floquet-Bloch theory applies and states that the field on the periodic lattice repeats itself with a propagation phase proportional to the propagation wave vector. Mathematically, for any angular frequency $\omega$,  a wave vector ${\bf k} (\omega) = k_x (\omega) \hat{x} + k_y (\omega) \hat{y}$ exists such that,
\begin{align}\label{eqn:bloch}
{\bf E} [i_x'p_x, i_y'p_y] =  P\left\{{\bf k}(\omega), \Delta i_x,  \Delta i_y \right\} {\bf E} [mp_x, np_y],
\end{align}
with the propagation phase defined as,
\begin{align}
 P\left\{{\bf k}(\omega), \Delta i_x,  \Delta i_y \right\} \equiv   e^{-j\left( k_x  \Delta i_x p_x + k_y  \Delta i_y p_y \right)}, \nonumber
\end{align}
where ${\bf E}$ is the field vector at each port defined on the edges of the unit cell in Fig. (\ref{fig:unit_cell}); $p_x, p_y$ are the lattice periods in $\hat{x}$ and $\hat{y}$ directions respectively; $(i_x, i_y)$ and $(i_x', i_y')$ are integers at two different lattice sites and $(\Delta i_x = i_x'-i_x, \Delta i_y = i_y'-i_y)$ is the vector difference between them. Note that due to periodicity, $k_{x/y} + M\pi/p_{x/y}, M=(-\infty, \infty)$ is considered to be equivalent. So, it is only necessary to consider $(k_x, k_y)$ in the first Brillouin Zone (BZ). Also, for the rest of this paper, we assume that the periods are the same for both $\hat{x}$ and $\hat{y}$ directions: $p_x = p_y =p$, following the $C_v$ symmetry \cite{sakoda_dirac_2012}.

It is well known that the transfer matrix connects the field at adjacent unit cells \cite{nada_theory_2017}, \cite{liao_optimal_2019} as follows,
\begin{align}\label{eqn:T}
    {\bf E} [(i_x+1)p, (i_y+1)p] = \overline{\overline{T}}  {\bf E} [mp, np].
\end{align}

Substituting Eq. (\ref{eqn:T}) into Eq. (\ref{eqn:bloch}), the following dispersion relation between the angular frequency $\omega$ and the wave vector ${\bf k}(\omega)$ as follows,
\begin{align}\label{eqn:dispersion} \left\{\overline{\overline{T}}(\omega) -  \overline{\overline{P}} \left\{{\bf k}(\omega), \Delta i_x,  \Delta i_y \right\}  \right\} {\bf E} [mp, np] =0, 
\end{align}
where $\overline{\overline{P}}$ is the phase propagation matrix depending on the wave vector ${\bf k}$ and the detailed arrangement of the field vector ${\bf E}$. The ${\bf k}-\omega$  dispersion curve is then obtained by solving the eigen-values/eigen-vectors problem of Eq. (\ref{eqn:dispersion}) after setting $\Delta i_x = \Delta i_y =1$ for the unit cell, which requires that the following secular equation to be satisfied,
\begin{align}\label{eqn:eigen} 
 \left[ \overline{\overline{T}}_{cell}(\omega) -  \overline{\overline{P}}_{cell}\left\{{\bf k}(\omega)\right\} \right]  \overline{i}  =0,
\end{align}
with $\overline{i}$ being the input array of ${\bf E} $ arranged in an appropriate sequence according to a specific geometry under study.

In order for the eigen problem of the secular equation in Eq. (\ref{eqn:eigen}) to have solution, its determinant should be zero,
\begin{align}\label{eqn:eigen_det} 
Det \equiv \left|\overline{\overline{T}}_{cell}(\omega) -  \overline{\overline{P}}_{cell}\left\{{\bf k}(\omega)\right\}  \right| =0.
\end{align}

It is clear that to obtain the dispersion relation according to Eq. (\ref{eqn:eigen}), the transfer matrix of the unit cell $\overline{\overline{T}}_{cell}(\omega)$ has to be obtained first. 

\section{Transfer Matrix of the 2D Interstitial Square Array}\label{sec:TMM_2D}
To obtain  the unit-cell transfer matrix $\overline{\overline{T}}_{cell}$ for the 2D interstitial square coupled MRRs array, the forward and backward propagating waves at each port of the unit cell have been defined in Fig. \ref{fig:unit_cell}: $a_i, i=1, 2, \cdots, 8$ are the forward propagating waves and $b_i, i=1, 2, \cdots, 8$ are the backward propagating waves at each port respectively.  The incident input waves array is then given as $\overline{i} = [{\bf a}_1, {\bf a}_2, {\bf a}_5, {\bf a}_6, {\bf b}_1, {\bf b}_2, {\bf b}_5, {\bf b}_6]'$, with the superscript $'$ being the transpose operator. Similarly, the output waves array is given by $\overline{o} = [{\bf b}_3, {\bf b}_4, {\bf b}_7, {\bf b}_8, {\bf a}_3, {\bf a}_4, {\bf a}_7, {\bf a}_8]'$. So the unit-cell transfer matrix is a 8-by-8 matrix that connects the input and output field vectors of the unit cell,
\begin{align}
 \overline{o} =   \overline{\overline{T}}_{cell}^{(8 \times 8)} \overline{i},
\end{align}
and the transfer matrix can be obtained by connecting all coupling ports with the $(4 \times 4)$ transfer matrices given in Table \ref{tab:matrices}. The $(4 \times 4)$  transfer matrices with various inputs combination can be derived from the $(4 \times 4)$  scattering matrix of the coupler, as shown in Appendix \ref{app:coupler_scattering}. Then the secular equation for the obtained $(8 \times 8)$  transfer matrix is obtained according to Eq. (\ref{eqn:eigen}),
\begin{align}\label{eqn:eigen_2D} 
\left\{\overline{\overline{T}}_{cell}^{(8 \times 8)}  - \begin{bmatrix}
 \overline{\overline{P}}_{x, cell}^{(4 \times 4)}\left\{k_x(\omega)\right\} & 0 \\
 0 &   \overline{\overline{P}}_{y, cell}^{(4 \times 4)}\left\{k_y (\omega)\right\}
\end{bmatrix} \right\}\overline{i} =0, 
\end{align}
with
\begin{align}
 \overline{\overline{P}}_{u, cell}^{(4 \times 4)}\left\{k_u(\omega)\right\} =  e^{-j\left( k_u(\omega)  p_u  \right)} \overline{\overline{I}}^{(4 \times 4)}, \ \ u=x, y. \nonumber
\end{align}

The eigen wave vector ${\bf k}(\omega) = [k_x(\omega), k_y(\omega)]'$  can be obtained by solving the determinant equation of Eq. (\ref{eqn:eigen_det}), 
\begin{align}\label{eqn:eigen_2D_det} 
\left|\overline{\overline{T}}_{cell}^{(8 \times 8)}  - \begin{bmatrix}
 \overline{\overline{P}}_{x, cell}^{(4 \times 4)}\left\{k_x(\omega)\right\} & 0 \\
 0 &   \overline{\overline{P}}_{y, cell}^{(4 \times 4)}\left\{k_y (\omega)\right\}
\end{bmatrix} \right| =0. 
\end{align}

\begin{table*}
\caption{\label{tab:matrices} $(4 \times 4)$  Transfer Matrices and Scattering Matrix of the Coupler} 
\begin{tabular}{ccccc}
Scattering Matrix &   Transfer Matrix: $(I, II)$ Input Ports &   Transfer Matrix: $(I, III)$ Input Ports\\
\begin{minipage}{4cm} \begin{equation} 
\overline{\overline{S}} =
\begin{bmatrix} 
0 & 0 & \tau & j \kappa  \\
0 & 0 & j \kappa &  \tau \\
\tau & j \kappa & 0 & 0   \\
 j \kappa & \tau & 0 & 0   
\end{bmatrix} \nonumber
\end{equation} \end{minipage}  & \begin{minipage}{4cm}\begin{equation}
\overline{\overline{T}}_{ (III,IV)}^{(I,II)} = 
\begin{bmatrix} 
j \kappa & \tau & 0 & 0  \\
\tau & j \kappa & 0 &  0 \\
0 &  0 & -j \kappa &  \tau   \\
0 &  0 & \tau & -j \kappa  
\end{bmatrix} \nonumber
\end{equation} \end{minipage}  & \begin{minipage}{4cm}\begin{equation}
\overline{\overline{T}}_{ (II,IV)}^{(I,III)}  = 
\begin{bmatrix} 
0 & j\frac{1}{\kappa} & -j\frac{\tau}{\kappa} & 0  \\
 j\frac{1}{\kappa}  &  0 & 0 &  -j\frac{\tau}{\kappa}   \\
j\frac{\tau}{\kappa} & 0 & 0 &  -j\frac{1}{\kappa} \\ 
0 &  j\frac{\tau}{\kappa} & - j\frac{1}{\kappa}  & 0 \kappa  
\end{bmatrix} \nonumber
\end{equation} \end{minipage}
\end{tabular} 
\end{table*} 

\section{Dispersion Relation} \label{sec:disp}

 The general form of the transfer matrix and the corresponding determinant of the secular equation given in Eq. (\ref{eqn:eigen}) are very complicated and contains little insight to understand the dispersion relation. Such general case can only be studied through numerical simulation, which is shown later. However, analytical results exist for special cases, which are of particular interest to understand some properties of the dispersion relation. Here the special  cases of equivalent arc phases or $\Phi_{L_i} \equiv \Phi_L \equiv mod(2 \pi L_i/\lambda, 2\pi), i =0, 1, 2, 3, 4$ are studied for: A) identical couplers, {\it i.e.}, $\kappa_i = \kappa_0 = \kappa, i = 1, 2, 3, 4$; and B) regular square coupled MRRs array,  {\it i.e.}, $\kappa_0 = 0$.
 
 \subsection{Identical Couplers}
In this case, analytical formulas are available for some typical arc lengths of the coupled MRRs and their phases defined as $\Phi_{L_i} \equiv \Phi_L \equiv 2 \pi L_i/\lambda, i =0, 1, 2, 3, 4$: i)  $1/4$-wavelength ring arc or $L_i = n/4 \lambda; \Phi_{L_i}=n/2\pi, n=1, 2, 3, \cdots$; ii)  $1/8$-wavelength ring arc or $L_i = n/8 \lambda; \Phi_{L_i}=n/4\pi, n=1, 2, 3, \cdots$.

\subsubsection{$1/4$-wavelength ring arc}
In this case, $\Phi_L = n\pi/2$ and the determinant of the secular equation is obtained as
\begin{align}\label{eqn:M_pi}
\left[P_x^2  a(k_y)  + P_x b(k_y) + c(k_y) \right]^2 = 0, 
\end{align}  
with
\begin{align}
& a(k_y) = P_y^2 + 2P_y (1-2\kappa^2) + 1, \nonumber \\
& b(k_y) = 2\left(P_y^2 +1\right) (1-2\kappa^2) + 4P_y(4\kappa^6 - 4\kappa^4 + 1), \nonumber \\
& c(k_y) =  P_y^2 + 2P_y (1-2\kappa^2) + 1,  \nonumber
\end{align}
where $P_x = e^{-j k_x p}$ and $P_y = e^{-j k_y p}$ are the propagation phases along $\hat{x}$ and $\hat{y}$ directions respectively.  

\begin{enumerate}[label=(\roman*)]
\item BZ line $\Gamma-M$ $(k_x, k_y = 0)$: 
In this case, $P_y =1$ and the determinant of the secular equation in Eq. (\ref{eqn:M_pi}) reduces to the following,
\begin{align}
\left[P_x^2  + 2P_x  \left(1- 2\kappa^4 \right)  +  1 \right]^2 = 0,
\end{align}  
from which we can obtain the four doubly-degenerated eigen values,
\begin{align}\label{eqn:1o4_lam_Gamma_M}
P_x^{(1, 2)} =  \left(2\kappa^4 -1\right) + j \sqrt{1 - \left(2\kappa^4 -1\right)^2}; \\
 P_x^{(3, 4)} =  \left(2\kappa^4 -1\right) - j \sqrt{1 - \left(2\kappa^4 -1\right)^2}.\nonumber
\end{align}

The wave vector $k_x$ can be obtained from Eq. (\ref{eqn:1o4_lam_Gamma_M}) as follows,
\begin{align}\label{eqn:kx_1o4_lam_Gamma_M}
k_x^{(1, 2)} = \frac{\arccos{ \left\{2\kappa^4 -1\right\}}}{p}; \\   k_x^{(3, 4)} = -\frac{\arccos{ \left\{2\kappa^4 -1\right\}}}{p} \nonumber,
\end{align}
and the eigen wave vectors $k_x^{(i)}, i=1, 2, 3, 4$ are real because $\left|2\kappa^4 -1\right| \le 1$ for $\kappa \in [0, 1]$.  

\item BZ line $M-X$ $(k_x, k_y = \pi/p)$:

In this case, Eq. (\ref{eqn:M_pi}) reduces to the following, 
\begin{align}
 \left[  P_x^2  - 2P_x  \left( 2\kappa^4 -  2\kappa^2 +1 \right)   + 1 \right]^2 = 0.
\end{align}   

Following similar procedure that leads to Eq. (\ref{eqn:kx_1o4_lam_Gamma_M}), the following is obtained,
\begin{align}\label{eqn:kx_1o4_lam_M_X}
k_x^{(1, 2)} = \frac{\arccos{ \left\{2\kappa^4 -  2\kappa^2 +1\right\}}}{p}; \\   k_x^{(3, 4)} = -\frac{\arccos{ \left\{2\kappa^4 -  2\kappa^2 +1\right\}}}{p} \nonumber,
\end{align}
and thus the eigen wave vectors $k_x^{(i)}, i=1, 2, 3, 4$ are also all real.


\end{enumerate}

\subsubsection{$1/8$-wavelength ring arc}\label{subsubsec:1o8_lambda}
In this case, $\Phi_L = n\pi/4$ and the determinant of the secular equation has the following coefficients,
\begin{align}
& a(k_y) = P_y^2 + 2P_y (\kappa^2-1) + 1, \nonumber \\
& b(k_y) = 2\left(P_y^2 +1\right) (2k^2-1) - 4P_y(4\kappa^6 - 12\kappa^4 + 8\kappa^2 - 1), \nonumber \\
& c(k_y) =  P_y^2 + 2P_y (2\kappa^2-1) + 1.  \nonumber
\end{align} 
  
\begin{enumerate}[label=(\roman*)]
\item BZ line $\Gamma-M$ $(k_x, k_y = 0)$: 
In this case, the determinant of the secular equation is obtained as
\begin{align}
 \left[P_x^2 -2 P_x\left(2\kappa^4 - 6\kappa^2 + 3 \right) + 1 \right]^2 = 0,
\end{align}  
from which the four eigen wave vectors $k_x^{(i)}, i=1, 2, 3, 4$  can be obtained as follows,
\begin{align}\label{eqn:kx_1o8_lam_Gamma_M}
k_x^{(1, 2)} = \frac{\hbox{arctanh} \left\{\sqrt{1 - \frac{1}{\left(2\kappa^4 - 6\kappa^2 + 3\right)^2}} \right\}}{p}; \\   k_x^{(3, 4)} = -\frac{\hbox{arctanh} \left\{\sqrt{1 - \frac{1}{\left(2\kappa^4 - 6\kappa^2 + 3\right)^2}} \right\}}{p} \nonumber,
\end{align}
which are imaginary for $\kappa \in \left[0, \left(3 - \sqrt{5}\right)/2 \right)$, which means exponential decay/loss and growth/gain for negative/positive imaginary values respectively. For $\kappa \in \left[\left(3 - \sqrt{5}\right)/2, 1\right]$, the eigen wave vectors $k_x^{(i)}, i=1, 2, 3, 4$  are real and given by,
\begin{align}\label{eqn:kx_1o8_lam_Gamma_M_real}
k_x^{(1, 2)} = \frac{\arccos{ \left\{2\kappa^4 - 6\kappa^2 + 3\right\}}}{p}; \\   k_x^{(3, 4)} = -\frac{\arccos{ \left\{2\kappa^4 - 6\kappa^2 + 3\right\}}}{p} \nonumber.
\end{align}

\item BZ line $M-X$ $(k_x, k_y = \pi/p)$: 

In this case, the determinant of the secular equation is obtained as follows,
\begin{align}
 \left[P_x^2 -2 P_x\left(2\kappa^4 - 4 \kappa^2 + 1 \right) + 1 \right]^2 = 0,
\end{align} 
from which the four doubly-degenerated eigen wave vectors $k_x^{(i)}, i=1, 2, 3, 4$  can be obtained as follows,
\begin{align}\label{eqn:kx_1o8_lam_M_X}
k_x^{(1, 2)} = \frac{\arccos{ \left\{2\kappa^4 - 4 \kappa^2 + 1\right\}}}{p}; \\   k_x^{(3, 4)} = -\frac{\arccos{ \left\{2\kappa^4 - 4 \kappa^2 + 1\right\}}}{p} \nonumber,
\end{align}
which are all real because $\left|2\kappa^4 - 4 \kappa^2 + 1\right| \le 1$.
\end{enumerate} 

\subsection{Regular Square Coupled MRRs Array}
When the interstitial rings are decoupled from the site rings, it becomes the regular square coupled MRRs array. The general dispersion can be obtained as follows, 
\begin{align}\label{eqn:4rings}
4\kappa^2\tau^2 \left(P_x^4 + 1\right) P_y^2 P_L^8  +  4\kappa^2 \tau^2 P_x^3P_y\left(P_y^2 + 1\right)P_L^4\left(P_L^8 + 1\right) + 4\kappa^2 \tau^2 P_x P_y\left(P_y^2 + 1\right)P_L^4\left(P_L^8 + 1\right) \ \ \ \ \ \  \ \ \ \ \ \  \ \ \ \ \ \  \ \ \ \ \ \  \ \ \ \ \ \ \ \ \\
  \ \ \ \ \ \  \ \ \ \     + P_x^2\left\{\left(1+P_y^4\right)\left( 4\kappa^2 \tau^2 P_L^8\right) + P_y^2\left(P_L^4 - 2\kappa P_L^2 + 1\right)\left(P_L^4 + 2\kappa P_L^2 + 1\right)\left[P_L^8 + 2(2\kappa^2 - 1)P_L^4 + 1\right]\right\}  = 0. \nonumber  
\end{align} 

\subsubsection{$1/4$-wavelength ring arc}
When the arc length of the ring is an integer number $n$ of quarter wavelength ($n/4\lambda$), $\Phi_L=n/2 \pi$ and Eq. (\ref{eqn:4rings}) reduces to the following,
\begin{align}
\left(P_x + P_y\right)^2 \left(P_xP_y + 1\right)^2 = 0,
\end{align}
which doesn't depend on the coupling coefficient $\kappa$ and two doubly degenerated solutions are obtained,
\begin{align}
   P_x = - P_y; \ \ \ \ \hbox{and} \ \ \ \    P_x = - \frac{1}{P_y},
\end{align}
and the doubly-degenerated eigen wave vectors are given by,
\begin{align}
k_x^{(1, 2)} = \frac{\pi}{p} + k_y, \ \ k_x^{(3, 4)} = \frac{\pi}{p} - k_y. 
\end{align}

In particular, the four eigen wave vectors are all degenerated for $k_y = 0$ or $k_y = \pi/p$,
\begin{align}\label{eqn:k0_kx_1o4}
k_x^{(1, 2, 3, 4)} = \frac{\pi}{p}, \ \ \hbox{for} \ \ k_y=0, \\
k_x^{(1, 2, 3, 4)} = 0, \ \ \hbox{for} \ \ k_y=\frac{\pi}{p}, \nonumber
\end{align}
where $k_x = 0$ and $k_x = 2\pi/p$ are equivalent due to the periodicity.

\subsubsection{$1/8$-wavelength ring arc}
When the arc length of the ring is an  integer number $n$ of 1/8 wavelength ($n/8 \lambda$), $\Phi_L=n/4\pi$ and Eq. (\ref{eqn:4rings}) reduces to the following,
\begin{align}
\left(P_x - P_y\right)^2\left(P_xP_y - 1\right)^2 = 0,
\end{align}
from which the two doubly degenerated solutions are obtained as follows,
\begin{align}
   P_x =  P_y; \ \ \ \ \hbox{and} \ \ \ \    P_x =  \frac{1}{P_y},
\end{align}
and the four doubly-degenerated eigen wave vectors given by,
\begin{align}
k_x^{(1, 2)} = k_y, \ \ k_x^{(3, 4)} = \frac{\pi}{p} - k_y. 
\end{align}

Similarly, the four eigen wave vectors are all degenerated for for $k_y = 0$ or $k_y = \pi/p$,
\begin{align}\label{eqn:k0_kx_1o8}
k_x^{(1, 2, 3, 4)} = 0, \ \ \hbox{for} \ \ k_y=0, \\
k_x^{(1, 2, 3, 4)} = \frac{\pi}{p}, \ \ \hbox{for} \ \ k_y=\frac{\pi}{p}. \nonumber
\end{align}

\section{Band Gaps}\label{sec:band_gap}
The band gap of the dispersion can be obtained by finding the minimum distance of the dispersion curve along the BZ lines $\Gamma-M-X-\Gamma$. The solutions of $P_L$ for $\Gamma/M/X$ points can be obtained by setting $(k_x=0, k_y=0)$, $(k_x=0, k_y=\pi/p)$, $(k_x=\pi/p, k_y=\pi/p)$ respectively in the  transfer matrix $\overline{\overline{T}}^{(8 \times 8)}$ of the secular equation in Eq. (\ref{eqn:eigen_2D}).

\subsection{Identical Couplers}
In this case, $\kappa_0 = \kappa_i$ and the $\Gamma/M/X$ points are obtained by solving Eq. (\ref{eqn:Gamma}), Eq. (\ref{eqn:M}) and Eq. (\ref{eqn:X}) below respectively,
\begin{align}\label{eqn:Gamma}
 P_L^8 + P_L^4\left(2 - 4\kappa^4\right) + 1 =0 ,
\end{align}
\begin{align}
P_L^{16} + P_L^{12}\left(16\kappa^6 - 40\kappa^4 + 24\kappa^2 - 4\right) + P_L^8\left(16\kappa^8 - 64\kappa^6 + 96\kappa^4 - 48\kappa^2 + 6\right) \nonumber \\
+ P_L^4\left(16\kappa^6 - 40\kappa^4 + 24\kappa^2 - 4\right) + 1 =0, \nonumber
\end{align}
\begin{align}\label{eqn:M}
P_L^{16} + P_L^8\left(-16\kappa^8 + 16\kappa^6 - 2\right) + 1 =0 ,
\end{align}
\begin{align}
 P_L^{16} + P_L^{12}\left(-16\kappa^6 + 32\kappa^4 - 16\kappa^2\right) + P_L^8\left(-16\kappa^8 + 48\kappa^6 - 64\kappa^4 + 32\kappa^2 - 2\right) \nonumber \\
 + P_L^4\left(-16\kappa^6 + 32\kappa^4 - 16\kappa^2\right) + 1 =0, \nonumber
\end{align}
\begin{align}\label{eqn:X}
P_L^8 + P_L^4\left(4\kappa^4 - 4\kappa^2 + 2\right) + 1 =0 ,
\end{align}
\begin{align}
P_L^{16} + P_L^{12}\left(16\kappa^6 - 24\kappa^4 + 16\kappa^2 - 4\right) + P_L^8\left(16\kappa^8 - 32\kappa^6 + 48\kappa^4 - 32\kappa^2 + 6\right) \nonumber \\
+ P_L^4\left(16\kappa^6 - 24\kappa^4 + 16\kappa^2 - 4\right) + 1 =0. \nonumber
\end{align} 
  
\subsection{Regular Square Coupled MRRs Array}
The $Gamma/M/X$ points of the regular square coupled MRRs array can be obtained in a similar way. It is found that the $\Gamma$ point and the $X$ point are identical, which satisfies the following equation, 
\begin{gather}\label{eqn:Gamma_decoupled}
 P_L  = e^{j \pm \frac{\pi}{4}}; e^{j \pm \frac{3\pi}{4}},  \\
P_L ^ 4 \pm 2P_L^2  \left(1 - 2  \kappa ^ 2\right) + 1  =0. \nonumber
\end{gather}

The $M$ point equation is given by,
\begin{gather}\label{eqn:M_decoupled}
P_L = \pm 1; \pm j, \\
P_L^8 +2 P_L^4\left(8\kappa^4 - 8\kappa^2 + 1\right) + 1  =0. \nonumber
\end{gather}

\section{Eigen Modes}\label{sec:eigen_modes}
The eigen values and eigen modes can be obtained by solving the secular equation in Eq. (\ref{eqn:eigen_2D}). The general forms of the eigen values and eigen modes can only be obtained through numerical simulation. However, for some typical cases of particular interest, analytical formulas can help to understand the behaviors of the 2D interstitial square coupled MRRs array. 

\subsection{Identical Couplers}
When $\kappa = \kappa_0$, analytical forms of the eigen values can be obtained for some particular cases.

\subsubsection{$1/4$-wavelength ring arc}
In this case, the Bloch eigen modes are doubly degenerated and their eigen values can be obtained. However, the analytical forms of the eigen vectors are very complicated and they are better obtained numerically. 

\begin{enumerate}[label=(\roman*)]
\item For $k_y=0$,
\begin{align}
P_x^\pm(k_x) =  2\kappa^4 \pm 2\kappa^2 
\sqrt{\kappa^4 - 1} - 1. \nonumber 
\end{align}

\item For $k_y=\pi/p$,
\begin{align}
 P_x^\pm(k_x) =  2\kappa^2(\kappa^2-1)+1 \pm 2k\sqrt{\kappa^6 - 2\kappa^4 + 2\kappa^2 - 1}. \nonumber 
\end{align}

\end{enumerate} 
 
\subsubsection{$1/8$-wavelength ring arc}
In this case, the eigen values are given as follows, 
\begin{enumerate}[label=(\roman*)]
\item For $k_y=0$,
\begin{align}
 P_x^\pm(k_x) =  2\kappa^2(\kappa^2 - 3) + 3 \pm 2\sqrt{ \kappa^8 - 6\kappa^6 + 12\kappa^4 - 9\kappa^2 + 2 }. \nonumber
\end{align}
 
\item For $k_y=\pi/p$,
\begin{align}
 P_x^\pm(k_x) = 2\kappa^2(\kappa^2 - 2) + 1 \pm 2\kappa \sqrt{ \kappa^6 - 4\kappa^4 + 5\kappa^2 - 2 }. \nonumber
\end{align}
\end{enumerate} 

\begin{table*}
\caption{\label{tab:eigen_modes} Bloch Eigen Modes $[a_1, b_1, a_2, b_2, a_5, b_5, a_6, b_6]$ of the Regular Square Coupled MRRs Array} 
\resizebox{\columnwidth}{!}{%
\begin{tabular}{ccccc}
 $P_L = \pm 1, (k_x, k_y)=(\pi/p, 0)$ $M$ point: Bloch mode \#1 & $[0, 1, 1, 0, \kappa( \kappa-j\tau), \tau(\tau-j\kappa), -\kappa(\kappa-j\tau), -\tau(\tau-j\kappa)]$  \\ 
 $P_L = \pm 1, (k_x, k_y)=(\pi/p, 0)$ $M$ point: Bloch mode \#2 & $[1, 1, 0, 0, 1-\kappa(\kappa-j\tau), \kappa(\kappa+j\tau), -1 +\kappa(\kappa-j\tau), -\kappa(\kappa+j\tau)]$  \\ 
 \hline
 $P_L = \pm 1, (k_x, k_y)=(0, \pi/p)$ $M$  point: Bloch mode \#1 & $[0, 1, -1, 0, -\kappa(\kappa+j\tau), -\tau(j\kappa+\tau), -\kappa(\kappa+j\tau), -\tau(j\kappa+\tau)]$  \\ 
 $P_L = \pm 1, (k_x, k_y)=(0, \pi/p)$ $M$ point: Bloch mode \#2 & $[-1, 0, 0, 1, -\kappa(\kappa+j\tau)+1, \kappa(\kappa-j\tau), -\kappa(\kappa+j\tau)+1, \kappa(\kappa-j\tau)]$  \\ 
  \hline
 $P_L = \pm j, (k_x, k_y)=(0, 0)$ $\Gamma$ point: Bloch mode \#1 & $[0, 1, -1, 0, -\kappa(j\kappa-\tau), -\tau(\kappa-j\tau), -\kappa(j\kappa-\tau), -\tau(\kappa-j\tau)]$  \\ 
 $P_L = \pm j, (k_x, k_y)=(0, 0)$ $\Gamma$  point: Bloch mode \#2 & $[-1, 0, 0, 1, \tau(\kappa+j\tau), -\kappa(j\kappa+\tau), \tau(\kappa+j\tau), -\kappa(j\kappa+\tau)]$  \\
  \hline
  $P_L = \pm j, (k_x, k_y)=(\pi/p, \pi/p)$ $X$ point: Bloch mode \#1 & $[0, 1, 1, 0, \kappa(j\kappa+\tau), -\tau(\kappa+j\tau), -\kappa(j\kappa+\tau), \tau(\kappa+j\tau)]$  \\ 
 $P_L = \pm j, (k_x, k_y)=(\pi/p, \pi/p)$ $X$ point: Bloch mode \#2 & $[1, 0, 0, 1, -\tau(\kappa-j\tau), -\kappa(j\kappa-\tau), \tau(\kappa-j\tau), \kappa(j\kappa-\tau)]$  
\end{tabular} %
}

\end{table*} 

\begin{figure}[t]
\includegraphics[width=1\linewidth]{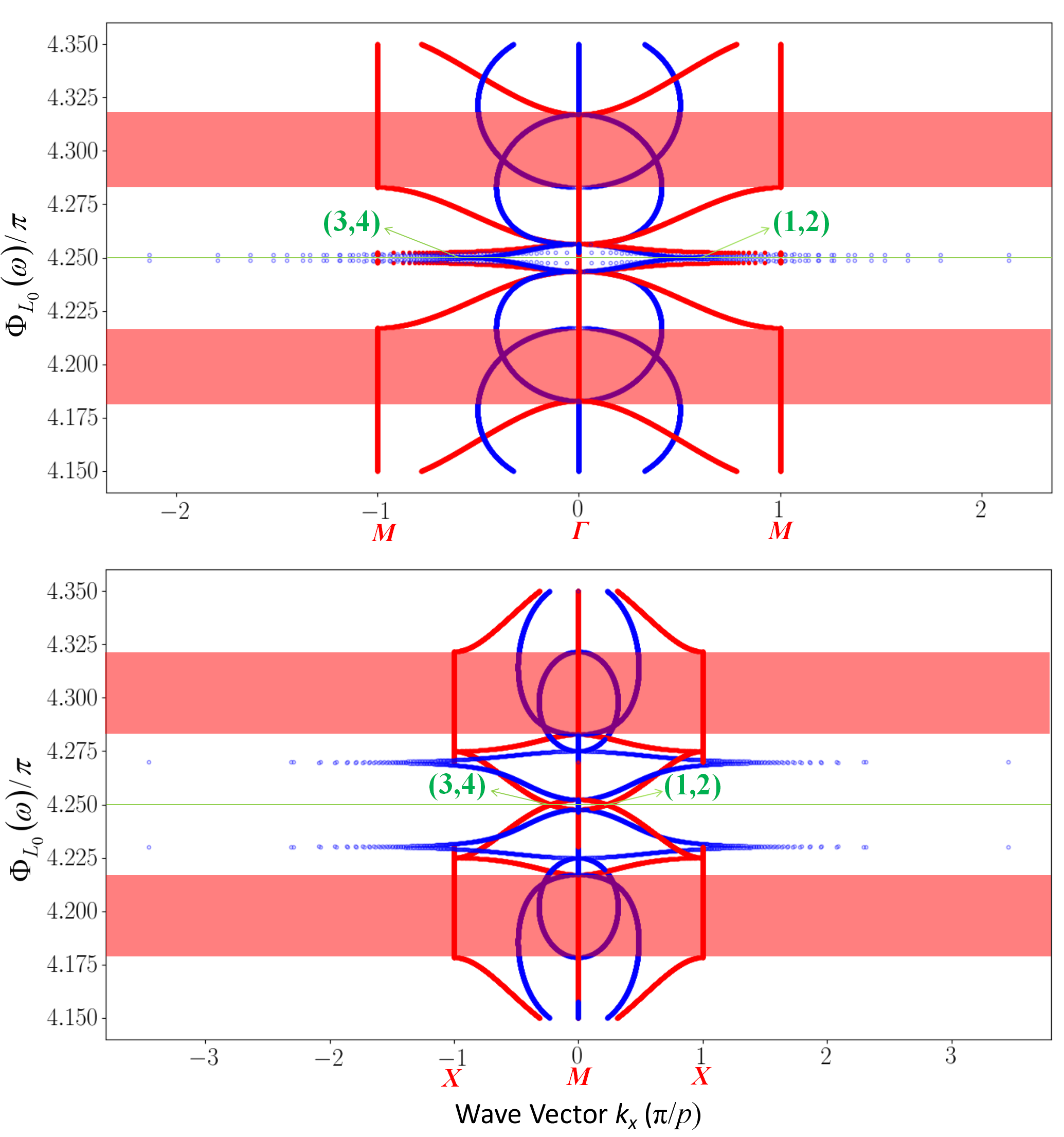}
\caption{\label{fig:dispersion_1o16} Dispersion relation for the 2D interstitial square coupled MRRs array of identical couplers ($\kappa_0 = \kappa = 1/4$) around $\Phi_{L_0} = 4.25 \pi$: top) along $\Gamma-M$ BZ line ($k_y =0$); and bottom) along $M-X$ BZ line ($k_y =\pi/p$).}
\end{figure}

\begin{figure}[t]
\includegraphics[width=1\linewidth]{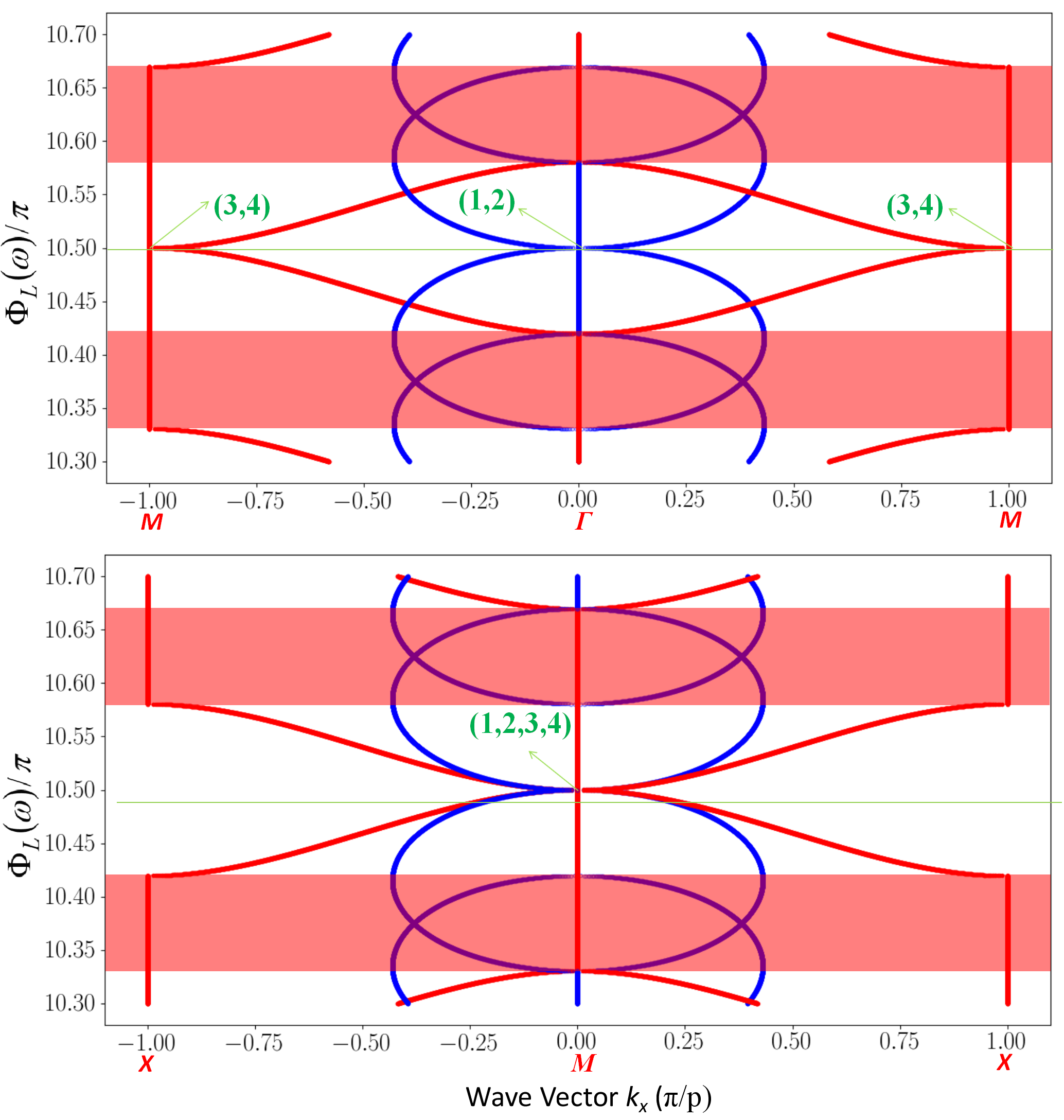}
\caption{\label{fig:dispersion_1o16_10p5_k0} Dispersion relation for regular the square coupled MRRs array ($\kappa_0 = 0$ and $\kappa = 1/4$) around $\Phi_{L} = 10.5 \pi$: top) along $\Gamma-M$ BZ line ($k_y =0$); bottom) along $M-X$ BZ line ($k_y =\pi/p$).}
\end{figure}
\subsection{Regular Square Coupled MRRs Array}
In this case, the eigen values/eigen modes for the $\Gamma/M/X$ points of some typical ring arc lengths can be obtained.

\subsubsection{$1/4$-wavelength ring arc}
In this case, $P_L = \pm 1$ and the four eigen modes reduce to two degenerated Bloch modes at the BZ $M$ points $(k_x, k_y) = (\pi/p, 0)$ and $(k_x, k_y) = (0, \pi/p)$. The Bloch modes (both eigen values and eigen vectors)  are given in Table \ref{tab:eigen_modes} for the incident input waves array  $\overline{\overline{i}} = [a_1, a_2, a_5, a_6, b_1, b_2, b_5, b_6]'$ and the output waves array $\overline{\overline{o}} = [b_3, b_4, b_7, b_8, a_3, a_4, a_7, a_8]'$ are related to the input waves through the Bloch wave condition with the corresponding eigen values according to the secular equation in Eq. (\ref{eqn:eigen_2D}).

\subsubsection{$1/8$-wavelength ring arc}
In this case, $P_L = \pm j$  and the 4 eigen modes reduce to two degenerated Bloch modes at the BZ $\Gamma$ point $(k_x, k_y) = (0, 0)$ and BZ $X$ point $(k_x, k_y) = (\pi/p, \pi/p)$. The Bloch modes are also given in Table \ref{tab:eigen_modes}.

\begin{figure*}[ht]
\includegraphics[width=1\linewidth]{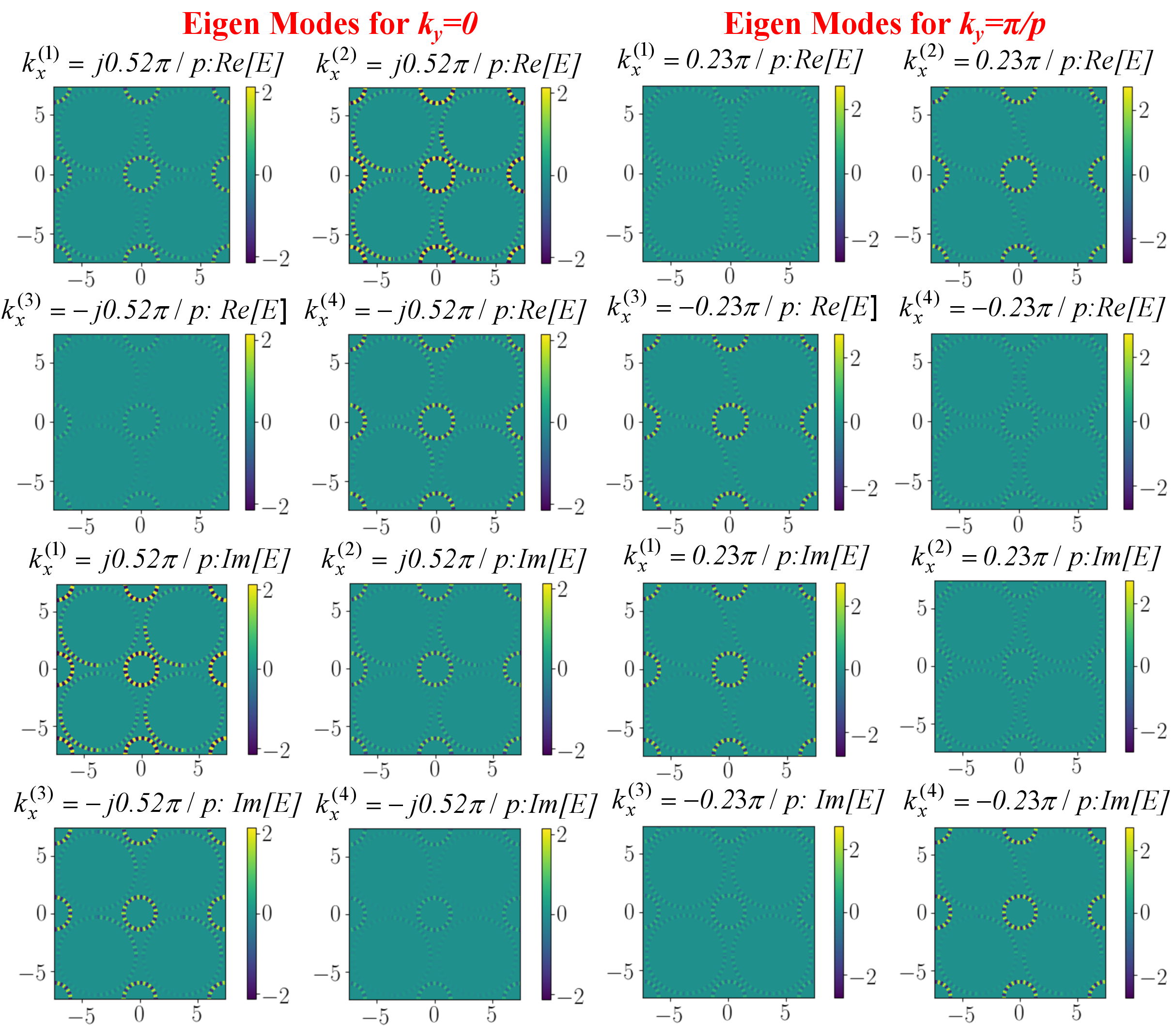}
\caption{\label{fig:field_4p25} Eigen modes' field distribution for the doubly-degenerated four modes in Fig. \ref{fig:dispersion_1o16}: $\kappa_0 = 0$ at $P_{L_0} = 4.25 \pi$, with the real/imaginary parts of the eigen modes' field distribution shown on top/bottom respectively: left) for $k_y = 0$ and the four doubly-degenerated eigen values are imaginary  with $k_x^{(1, 2)} =j 0.52\pi/p$ (gain)  and  $k_x^{(3, 4)} =-j 0.52\pi/p$ (loss); right) $k_y = \pi/p$ and all the four doubly degenerated eigen values are all real (propagating modes) with $k_x^{(1, 2)} =0.23 \pi/p$ (gain)  and  $k_x^{(3, 4)} =-0.23 \pi/p$.}
\end{figure*}

\begin{figure*}[ht]
\includegraphics[width=1\linewidth]{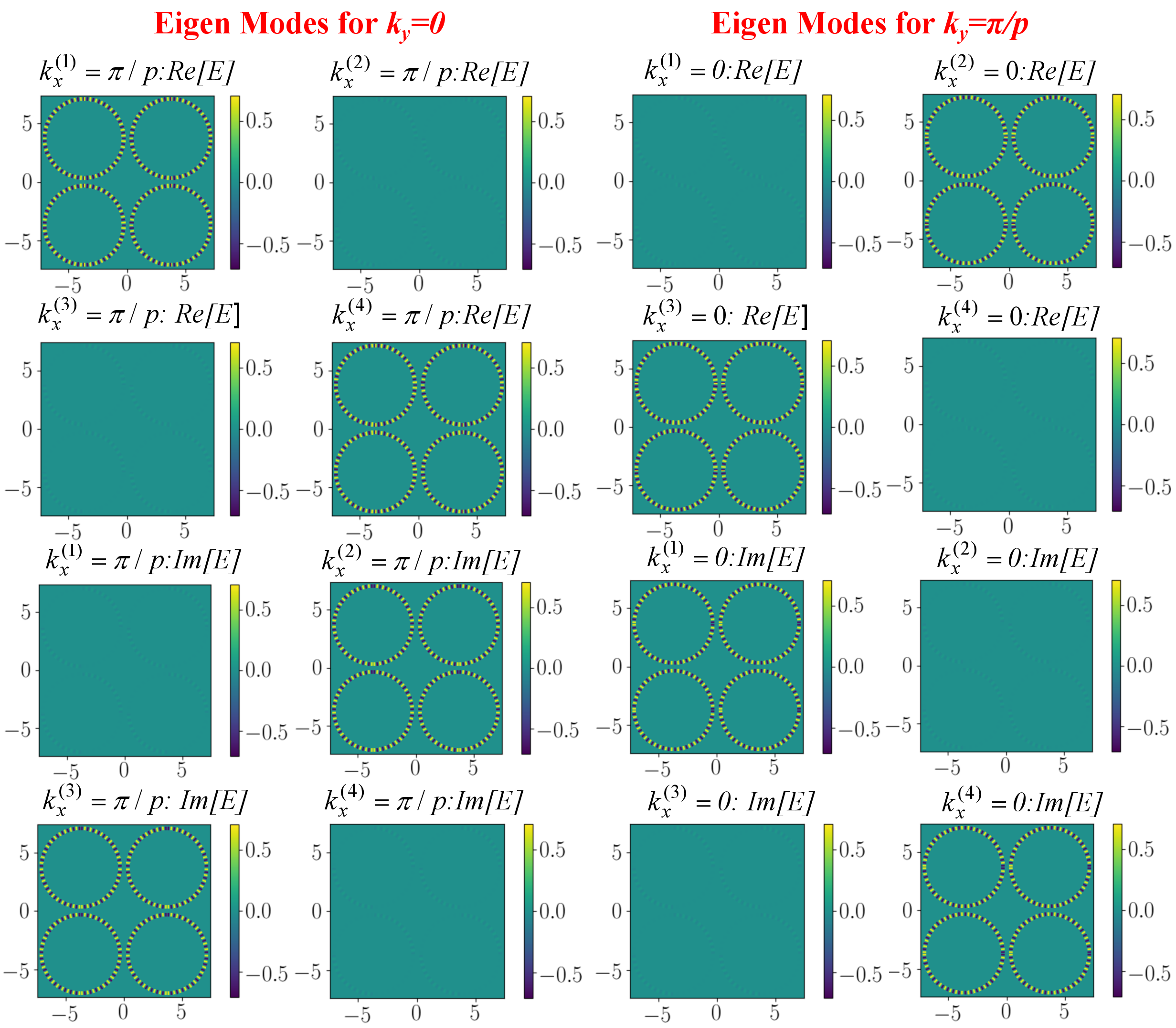}
\caption{\label{fig:field_10p5} Eigen modes' field distribution for the four doubly-degenerated modes for the regular square coupled MRRs array ($\kappa_0 = 0$) at $\Phi_{L} = 10.5 \pi$, with the real/imaginary parts of the eigen modes' field distribution shown on top/bottom respectively: left) $k_y = 0$ with the four degenerated eigen wave vectors of $k_x^{(1, 2, 3, 4)} = \pi/p$; and right) $k_y = \pi/p$ with the four degenerated eigen wave vectors of $k_x^{(1, 2, 3, 4)} =0$.}
\end{figure*}

\begin{figure*}[ht]
\includegraphics[width=0.95\linewidth]{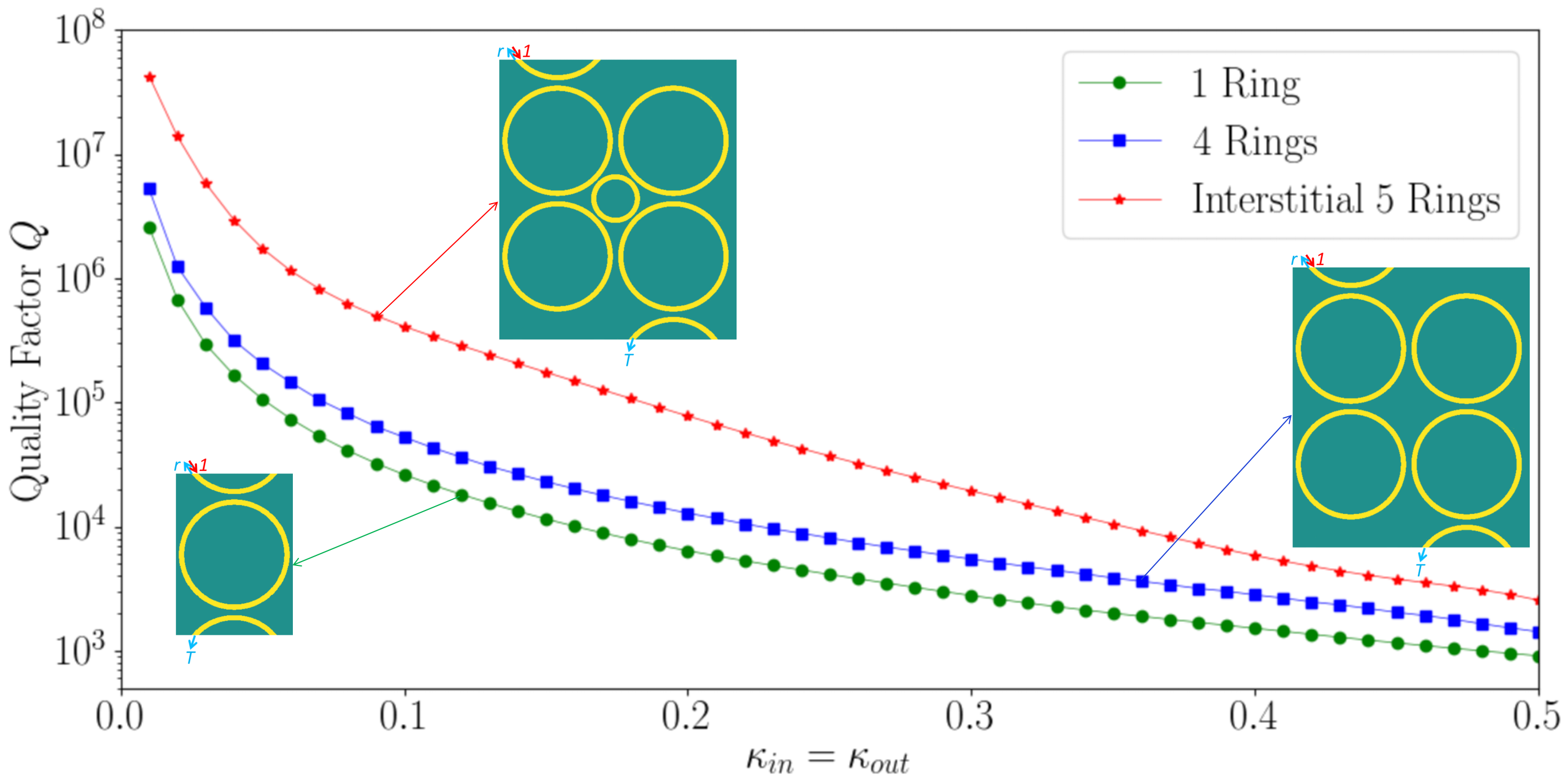}
\caption{\label{fig:Q} Loaded quality factors $Q$ vs. the input/output coupling coefficients $\kappa_{in} = \kappa_{out}$ of three MRRs configurations: i) 1-ring configuration (green dots); ii) 4-ring configuration (blue squares); and iii) interstitial 5-ring configuration (red stars).}
\end{figure*}

\section{Simulation Results} \label{sec:results}
Numerical simulation has been performed to obtain the dispersion curves and eigen modes for the 2D interstitial square coupled MRRs array. Here results are shown for some typical cases: 1) identical couplers with a  $\lambda/8$ interstitial-ring arc; and 2) regular square coupled MRRs array with a $\lambda/4$ site-ring arc. During simulation a coupling strength of $\kappa=1/4$ is used. 

\subsection{Dispersion Curves}
During the simulation, two BZ lines are calculated: i) $\Gamma-M$ BZ line; and 2) $M-X$ BZ line,  by setting $k_y =0$ and $k_y = \pi/p$ respectively. When $k_y$ is given, the $(8 \times 8)$ transfer matrix in Eq. (\ref{eqn:eigen_2D}) reduces to a $(4 \times 4)$ matrix with the variable of $k_x$, which gives four eigenvalues and their corresponding eigen modes for an angular frequency $\omega$. 



\subsubsection{Identical Couplers}
In this case, a 2D interstitial square coupled MRRs array with identical couplers ($\kappa_0 = \kappa = 1/4$) is studied. The arc length of the site-ring is $L_i = 5 \frac{1}{8} \lambda, i=1, 2, 3, 4$, giving a radius of $R_i = 4 L_i/\pi \sim 10 \mu m$; and the arc length of the interstitial-ring is $L_0 = 2 \frac{1}{8} \lambda$, giving a radius of $R_0 = 4 L_0/\pi \sim 4 \mu m$. 

The top plot of Fig. \ref{fig:dispersion_1o16} shows the result for the $\Gamma-M$ BZ line when $k_y =0$: the red/blue lines are the real/imaginary parts of the four eigen values of $k_x^{(i)}, i=1, 2, 3, 4$ and the red shaded areas are the bad gaps. Note that the wave vector $k_x$ of the horizontal axis is in the unit of $\pi/p$ and the vertical axis is the phase of the interstitial-ring's arc length $\Phi_{L_0} = 2 \pi L_0/\lambda$ in the unit of $\pi$. From Fig. \ref{fig:dispersion_1o16}, at $\Phi_{L_0} = 4 \frac{1}{4} \pi$, the four modes degenerates to two doubly-degenerated eigen modes, as expected in the analytical result given in Eq. (\ref{eqn:kx_1o8_lam_Gamma_M}). The one of the two doubly-degenerated eigen modes is a mode with loss and the other one is a mode with gain because $\kappa = 1/4  \in \left[0, \left(3 - \sqrt{5}\right)/2 \right)$. The eigen wave vectors are obtained as $k_x^{(1,2)} = j 0.52 \pi/p$ and $k_x^{(3,4)} = -j 0.52\pi/p$, verifying the analytical result given in Eq. (\ref{eqn:kx_1o8_lam_Gamma_M}).

Also, the bottom plot of Fig. \ref{fig:dispersion_1o16} shows the result for the $M-X$ BZ line when $k_y =\pi/p$: the four doubly-degenerated eigen wave vectors are obtained as $k_x^{(1,2)} = 0.23\pi/p$ and $k_x^{(3,4)} = -0.23\pi/p$, also verifying the analytical result given in Eq. (\ref{eqn:kx_1o8_lam_M_X}).

At last, the band gaps (red shaded areas) can be verified according to Eq. (\ref{eqn:Gamma}) to Eq. (\ref{eqn:X}).





\subsubsection{Regular Square Coupled MRRs Array}

In this case, a decoupled regular square coupled MRRs array with a coupling strength of $\kappa = 1/4$ is studied. The arc length of the site-ring is $L_i = 5 \frac{1}{8} \lambda, i=1, 2, 3, 4$.

The top plot of Fig. \ref{fig:dispersion_1o16_10p5_k0} shows the dispersion curve along the $\Gamma-M$ BZ line ($k_y =0$), from which the eigen wave vectors at $\Phi_L = 10 \frac{1}{2} \pi$ are obtained as four degenerated eigen wave vectors: $k_x^{(1, 2, 3, 4)} = \pi/p$, verifying the analytical result given in Eq. (\ref{eqn:k0_kx_1o4}). 

Similarly, the bottom plot of. \ref{fig:dispersion_1o16_10p5_k0}  shows the dispersion curve along the $M-X$ BZ line ($k_y =\pi/p$), from which the eigen wave vectors at $\Phi_L = 10 \frac{1}{2} \pi$ are obtained as four degenerated eigen wave vectors: $k_x^{(1, 2, 3, 4)} = 0$, verifying the analytical result given in Eq. (\ref{eqn:k0_kx_1o8}).

At last, the band gaps (red shaded areas) can be verified according to Eq. (\ref{eqn:Gamma_decoupled}) and Eq. (\ref{eqn:M_decoupled}).




\subsection{Eigen Modes}
After the eigen wave vectors are obtained, the eigen modes' field distribution can also be calculated through the secular equation given in Eq. (\ref{eqn:eigen_2D}). 

\subsubsection{Identical Couplers}
The left plots of Fig. \ref{fig:field_4p25} show the four doubly-degenerated eigen modes' field distribution at $k_y = 0$, with the  corresponding eigen wave vectors $k_x^{(1,2)} = j 0.52\pi/p$ (mode with gain) and $k_x^{(3,4)} = -j 0.52\pi/p$ (mode with loss) shown in the top dispersion plot of Fig. \ref{fig:dispersion_1o16}. The real/imaginary parts of the eigen modes' field are shown on the top/bottom plots respectively.

Also, the right plots of Fig. \ref{fig:field_4p25} show the eigen modes' field distribution at $k_y = \pi/p$, with the corresponding eigen wave vector  $k_x^{(1,2)} = 0.23\pi/p$ (mode propagating forward) and $k_x^{(3,4)} = - 0.23\pi/p$ (mode propagating backward) shown in the bottom dispersion plot of Fig. \ref{fig:dispersion_1o16}.

\subsubsection{Regular Square Coupled MRRs Array}
Similarly, Fig. \ref{fig:field_10p5} show the eigen modes' field distribution for the regular square coupled MRRs array ($\kappa_0 =0$): left) $k_y =0$ with a four degenerated eigen wave vector of $k_x^{(1, 2, 3, 4)} = \pi/p$ (BZ M point: $k_x=\pi/p, k_y=0$), corresponding to the top dispersion plot in Fig. \ref{fig:dispersion_1o16_10p5_k0}; and right) $k_y = \pi/p$ with a four degenerated eigen wave vector of $k_x^{(1, 2, 3, 4)} = 0$ (BZ M point: $k_x=0, k_y=\pi/p$), corresponding to the bottom dispersion plot in Fig. \ref{fig:dispersion_1o16_10p5_k0}. Due to the $C_{4 v}$ symmetry of the regular square coupled MRRs array \cite{sakoda_dirac_2012}, the four degenerated eigen modes of   $k_y = \pi/p$ can be obtained by rotating those of $k_y =0$ by $90^\circ$.


\section{Applications}\label{sec:application}
MRRs and arrays have the great potential applications in areas such as integrated filters and resonance based sensing. One important common parameter of MRRs is the quality factor $Q$, which is related to the energy dissipation rate of the MRRs. The total quality factor includes both the quality factor due to the intrinsic loss of the MRRs, {\it i.e.}, $Q_0$ and the loaded quality factor due to the coupling to the input/output couplers, {\it i.e.}, $Q_L$. The quality factor can be studied in two ways \cite{nada_theory_2017}: 1) through the group velocity according to the dispersion relation $Q \propto 1/\left|{\bf v}_g(\omega)\right| = 1/\left|\nabla_{\bf k} \omega\right|$; and 2) through the Full Width at Half Maximum (FWHM) as $Q = 1/\hbox{FWHM}$. The group velocity method of 1) can provide qualitative insight about the $Q$ through the dispersion curves such as those in Fig. \ref{fig:dispersion_1o16} and Fig. \ref{fig:dispersion_1o16_10p5_k0}; While the FWHM method of 2) is a convenient numerical way to calculate the $Q$, with the help of the transfer matrix of the unit cell of the interstitial square coupled MRRs array in Eq. (\ref{eqn:eigen_2D}) and appropriate terminating boundary conditions. To do this, the reflection $r$ and transmission $T$ in Fig. \ref{fig:Q} can be expanded into the superposition of the eigen modes such as those in Fig. \ref{fig:field_4p25} and Fig. \ref{fig:field_10p5} with unknown coefficients; then the solution  can be obtained by imposing the corresponding boundary conditions. Here the loaded quality factors $Q_L$ for three MRRs configurations  are calculated with the FWHM method of 2) and compared: i) the 1-ring configuration; ii) the 4-ring configuration; and iii) the interstitial 5-ring configuration. Fig. \ref{fig:Q} shows the obtained loaded quality factor $Q_L$ vs. the input/output coupling coefficients $\kappa_{in} = \kappa_{out} \in(0, 0.5]$ of the input/output couplers at the top/bottom respectively, for a ring-ring coupling coefficient of $\kappa_i = \kappa_0 = 1/4, i=1, 2, 3, 4$. It can be seen that the interstitial 5-ring configuration has a loaded quality factor up to 20 times as high as that of the 1-ring configuration, {\it i.e.}, $Q_L^{(5)}$ = 20 $Q_L^{(1)}$, and up to 8 times as high as that of the 4-ring configuration,  {\it i.e.},  $Q_L^{(5)}$  = 8 $Q_L^{(4)}$. This is partly due to that more rings increases the energy storage length. But more importantly, it is due to the dispersion enhancement of the structure geometry configuration: the degeneration of the eigen modes such as those in Fig. \ref{fig:dispersion_1o16} allows mixing of the degenerated eigen modes by the incident wave at the input port, greatly increasing the energy stored per unit length of the MRRs and the loaded quality factor is thus dramatically increased. Such high value of loaded quality factor of the interstitial coupled MRRs array makes it a strong candidate to form high-quality filters and resonance based sensing devices. Also, note that the quality factor due to the intrinsic loss $Q_{loss}$ can be readily calculated and the total quality factor is given by $1/Q = 1/Q_L + 1/Q_{loss}$. What's more important, when both loss and gain are introduced and balance each other to form the parity-time symmetric sensor, higher quality factor can be obtained \cite{Liao_PT_IEEE_TAP_2019}. Finally, the quality factor can be further improved by tuning the coupling coefficient of $k_i, i=0, 1, 2, 3, 4$ and $\kappa_0$ individually, providing more flexibility to design better integrated photonics components with the interstitial square coupled MRRs array.

\section{Conclusion}\label{sec:con}
In this paper, the interstitial square coupled MRRs array has been investigated. TMM is used to obtain the dispersion relation under the Floquet-Bloch periodic condition. Analytical formulas of eigen wave vectors, band gaps and eigen modes' field distribution are derived for the particular cases of the interstitial square coupled MRRs array with identical couplers and the regular square coupled MRRs array. It is found that the eigen modes are doubly degenerated for the 2D interstitial square coupled MRRs array  and all four eigen modes are degenerated for the regular 2D square coupled MRRs array,  when the arc length is multiple integer of quarter wavelength. Then, the eigen mode field distribution is calculated through the secular equation after the four eigen wave vectors along a BZ line for a given frequency is obtained.  Numerical simulation is performed for both a 2D interstitial coupled MRRs array with identical couplers and a regular square coupled MRRs array; the simulation results verify the analytical analysis. At last, it is found that the loaded quality factors of the interstitial 5-ring configuration is up to 20 times and 8 times as high as those of the 1-ring configuration and the regular 4-ring configuration respectively. Such high quality factor indicates that the interstitial square coupled MRRs array has the potential to form high-quality integrated photonics components such as filters and resonance based sensing devices.


\appendix

\begin{figure}[h]
\centering
\includegraphics[width=0.75\linewidth]{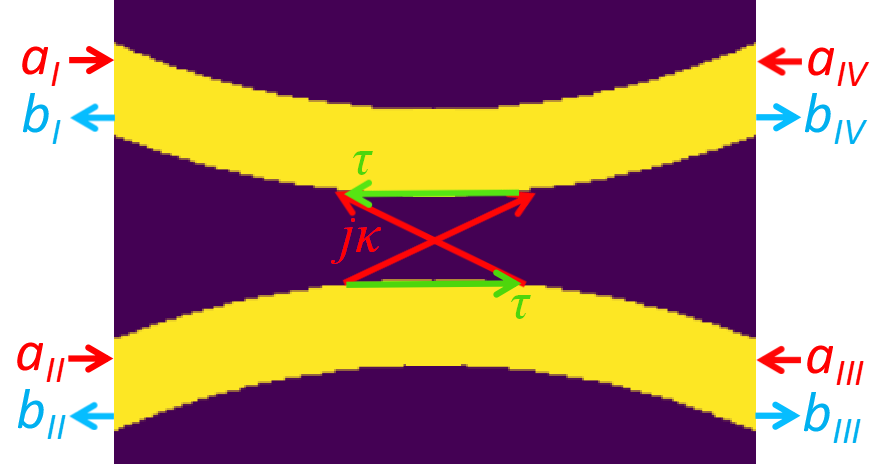}
\caption{\label{fig:coupler} Ports definition of the coupler.}
\end{figure}

\section{ $(4 \times 4)$ Scattering and Transfer Matrices  of the Coupler}\label{app:coupler_scattering}
The definition of each port of the $(4 \times 4)$ coupler is shown in Fig. \ref{fig:coupler}. The incident input waves array and outgoing waves array are given by $\overline{a} \equiv [a_I, a_{II}, a_{III}, a_{IV}]'$ and $\overline{b} \equiv [b_I, b_{II}, b_{III}, b_{IV}]'$ respectively.
\begin{equation}\label{eqn:coupler_scattering}
\overline{b} = \overline{\overline{S}}^{(4 \times 4)} \overline{a}:  \ \ \ \  \overline{\overline{S}}^{(4 \times 4)} = 
\begin{bmatrix} 
0 & 0 & \tau & j \kappa  \\
0 & 0 & j \kappa &  \tau \\
\tau & j \kappa & 0 & 0   \\
 j \kappa & \tau & 0 & 0   
\end{bmatrix}
\end{equation}
 
Depending on the input ports and output ports, the transfer matrix can take different forms. The transfer matrix can be obtained by re-arranging the input/output waves in the scattering matrix given in Eq. (\ref{eqn:coupler_scattering}). The following two transfer matrices are used for the 2D interstitial square coupled MRRs array.
\begin{enumerate}[label=(\roman*)]
\item $(I, II)$ Input ports/$(III, IV)$ Output ports: 
\begin{equation}\label{eqn:coupler_transfer_1234}
\overline{o}_{(III, IV)} = \overline{\overline{T}}_{ (III,IV)}^{(I,II)}  \overline{i}_{(I, II)},
\end{equation}
where $\overline{i}_{(I, II)} = [a_I, a_{II}, b_{I}, b_{II}]'$ and  $\overline{o}_{(III, IV)} = [a_{III},  a_{IV}, b_{III}, b_{IV}]'$; and the transfer matrix $\overline{\overline{T}}_{ (III,IV)}^{(I,II)}$ is given in Table \ref{tab:matrices}.

\item $(I, III)$ Input ports/$(II, IV)$ Output ports: 
\begin{equation}\label{eqn:coupler_transfer_1324}
\overline{o}_{(II, IV)} = \overline{\overline{T}}_{ (II,IV)}^{(I,III)}  \overline{i}_{(I, III)},
\end{equation}
where $\overline{i}_{(I, III)} = [a_I, a_{III},  b_I,  b_{III}]'$ and  $\overline{o}_{(II, IV)} = [a_{II},  a_{IV}, b_{II}, b_{IV}]'$; and the transfer matrix $\overline{\overline{T}}_{ (II,IV)}^{(I,III)}$ is given in Table \ref{tab:matrices}.

\end{enumerate} 

\section{$(8 \times 8)$ Transfer Matrix}\label{app:coupling_equations}
The $(8 \times 8)$ transfer matrix $\overline{\overline{T}}_{cell}^{(8 \times 8)}$  in Eq. (\ref{eqn:eigen_2D}) can be obtained by connecting all the input ports to all the output ports through the two $(4 \times 4)$ transfer matrices $\overline{\overline{T}}_{ (III,IV)}^{(I,II)}$ and $\overline{\overline{T}}_{ (II,IV)}^{(I,III)}$ given in Eq. (\ref{eqn:coupler_transfer_1234}) and Eq. (\ref{eqn:coupler_transfer_1324}) respectively. Here, ports $(1, 2, 5, 6)$ are used as the input ports with the input waves defined as $i_{1256} \equiv [a_1, a_2,  a_5, a_6, b_1, b_2, b_5, b_6]'$ and ports $(3, 4, 7, 8)$ are used as the output ports with the output waves defined as $i_{3478} \equiv [a_3, a_4,  a_7, a_8, b_3, b_4, b_7, b_8]'$. Then $\overline{\overline{T}}_{ (II,IV)}^{(I,III)}$ is used for couplers $C_2$ and $C_4$ and $\overline{\overline{T}}_{ (III,IV)}^{(I,II)}$ is used for all other couplers to connect all input ports  $(1, 2, 5, 6)$ to all output ports $(3, 4, 7, 8)$.

\end{document}